%% file: manuscript.tex
\documentclass[]{interact}

\usepackage{epstopdf}
\usepackage[caption=false]{subfig}

\usepackage[utf8]{inputenc}
\usepackage{xcolor}
\usepackage[T1]{fontenc}
\usepackage{lmodern}
\usepackage{graphicx}
\usepackage{amsmath}
\usepackage{adjustbox}
\usepackage{tikz}
\usepackage{enumitem}
\usepackage{multirow}
\usepackage{booktabs}
\usepackage{comment}
\usepackage{color}
\usepackage{colortbl}
\usepackage[normalem]{ulem}

\usepackage{newtxtext,newtxmath}
\usepackage[colorlinks=true,citecolor=black,linkcolor=black]{hyperref}

\usepackage{nomencl}
\makenomenclature

\hypersetup{
    colorlinks = true,
    urlcolor   = black,
    citecolor  = black,
    linkcolor= black,
}

\usepackage[natbibapa,nodoi]{apacite}
\theoremstyle{plain}

\theoremstyle{definition}

\theoremstyle{remark}

\begin{document}
\articletype{PREPRINT \\
\lowercase{accepted for publication in the \uppercase{J}ournal of \uppercase{H}ydraulic \uppercase{R}esearch}}

\title{On the estimation of specific interfacial areas in aerated flows}

\author{
\name{M. Kramer\textsuperscript{a}\thanks{CONTACT M. Kramer. Email: m.kramer@unsw.edu.au}}
\affil{\textsuperscript{a}Senior Lecturer, UNSW Canberra, School of Engineering and Technology (SET), Canberra,
ACT 2610, Australia,  ORCID 0000-0001-5673-2751}
}

\maketitle

\begin{abstract}
This work provides a short clarification on the definition of specific interfacial areas in aerated flows, thereby distinguishing between the specific interfacial area related to the volume of an air-water mixture, and the specific interfacial area related to the water volume. This distinction is important when it comes to solving the air-water mass transfer equation, as there has been some misconception on the past. It is hoped that this contribution will help to clarify on these basic fundamentals, which is anticipated to be important for the future development and application of advanced air-water mass transfer models to aerated flows.
\end{abstract}

\begin{keywords}
Air-water flows; specific interfacial area; phase-detection probe
\end{keywords}

\section{Introduction}

In high Froude-number aerated flows, such as flows down smooth chutes or stepped cascades, the gas-liquid interfacial area is of key interest, as the mass transfer between air and water is directly proportional to this area. In the mass transfer equation, the interfacial area typically appears in form of a specific interfacial area, which represents an interfacial area per unit volume. In this work, some considerations on the definition and estimation of specific interfacial areas in aerated flows are presented. These considerations are deemed very relevant to several previous works, where aeration efficiencies, \textcolor{black}{or other relevant parameters to the mass transfer process},  were estimated through a combination of the mass transfer equation with phase-detection intrusive measurements.

In these previous works, the specific interfacial area was defined as the interfacial area per unit volume of air and water \citep{Gulliver1990,Toombes2002,Toombes2005,Wilhelms01092005,Bung2009,Felder2015a,Wuthrich2015,Bung2018,Severi2018,Niu2024,KC02012025}, and was computed on the basis of sidewall images or using phase-detection intrusive probe measurements. For the latter, the commonly adopted expression reads \citep{Cartellier1991,Chanson2002}    
\begin{equation}
a_\text{m} =   \frac{4 \, F }{V},
\label{Eq1}
\end{equation}
where $a_\text{m}$ is the specific interfacial area per unit volume of air and water, $F$ is the interfacial frequency, defined as half the rate of detected interfaces, $V$ is the time-averaged interfacial velocity, and the index m stands for mixture.

However, in the mass transfer equation, the interfacial area must be related to the volume of water  \citep{Wilhelms1993,Cussler2009}, and as such, the correct equation for the specific interfacial area reads
\begin{equation}
a_\text{w} =  \frac{4 \, F }{V \left(1 - C \right)},
\label{Eq1a}
\end{equation}
where $a_\text{w}$ is the specific interfacial area per unit volume of water, $C$ is the time-averaged volumetric air concentration, and the index w stands for water. 

The reasoning as to why the interfacial area must be related to the volume of water, as well as a derivation of $a_\text{w}$, is given in the next section, while the re-analysis of a literature data set shows that aeration efficiencies, estimated using $a_\text{m}$ instead of $a_\text{w}$, may be subject to significant errors.

\section{Methods}
\label{secDer}
\subsection{Air-water mass transfer equation}
Here, some considerations on mass transfer are presented, followed by a derivation of the expression for the 
interfacial area per unit volume of water. First, we are interested in the transfer of mass from an air-water interface into a well-mixed solution. The amount of transferred mass per interfacial area can be written in familiar form \citep{Cussler2009}
\begin{equation}
N = K_L \left(c_{\text{sat}} - c_{\text{gas}} \right),   
\end{equation}
where $N$ is the mass flux through the interface, $K_L$ is the mass transfer coefficient, $c_{\text{gas}}$ 
is the dissolved gas
concentration in the bulk solution, and $c_{\text{sat}}$ is the equilibrium gas concentration at the interface. Next, we write the rate \textcolor{black}{of} mass transfer between the interface and the water phase as follows
\begin{equation}
\underbrace{\frac{\text{d}  \left(c_{\text{gas}} \, \mathcal{V}_\text{w} \right)}{\text{d}t}}_{\substack{\text{Accumulation of mass} \\ \text{in water phase (kg/s)}}} =  N A = \underbrace{ K_L A \left(c_{\text{sat}} - c_{\text{gas}} \right) \vphantom{\frac{\text{d}  \left(\mathcal{V}_\text{w} \, c_{\text{gas}} \right)}{\text{d}t}}}_{\substack{\text{Rate of mass transfer} \\ \text{across interface}}},  
\label{Eq2}
\end{equation}
where $\mathcal{V}_w$ is the water volume, and $A$ is the interfacial area. We recognise that the water volume does not change over time, and as such, Eq. (\ref{Eq2}) is divided by $\mathcal{V}_w$ to yield
\begin{equation}
\frac{\text{d}   c_{\text{gas}} }{\text{d}t}  =  K_L \, \frac{A}{\mathcal{V}_w} \, \left(c_{\text{sat}} - c_{\text{gas}} \right) =  K_L \, a_\text{w} \left(c_{\text{sat}} - c_{\text{gas}} \right), 
\label{EqGas}
\end{equation}
where the specific interfacial area $a_\text{w} = A/\mathcal{V}_\text{w}$ is defined as the interfacial area per unit of water volume. Using $\text{d}s = V \text{d}t$, the following formulation is obtained
\begin{equation}
\frac{\text{d}   c_{\text{gas}} }{\text{d}s}  =   \frac{K_L \, a_\text{w}}{V} \left(c_{\text{sat}} - c_{\text{gas}} \right), 
\label{EqGas1}
\end{equation}
where $s$ is the distance along a streamline. Equation (\ref{EqGas1}) can be integrated to calculate the aeration efficiency of a hydraulic structure, given that $K_L$, $V$, and $a_\text{w}$ are known at each location \citep{Toombes2002}. 

\textcolor{black}{It is emphasized that water volume $\mathcal{V}_\text{w}$ and not the mixture volume $\mathcal{V}_\text{m}$ appears in Eq. (\ref{Eq2}), which can be explained using a control volume approach, where the interface of an air bubble is considered a control surface. As the mass transferred across the air bubble interface accumulates in the surrounding water phase, the rate of change of mass within the water phase must be expressed using the water volume, which subsequently leads to the definition of the specific interfacial area as $a_\text{w}=A/\mathcal{V}_\text{w}$.}

\subsection{Specific interfacial area}
Considering a bubbly flow with $n$ uniform-sized bubbles, each bubble having a surface area $A_b = \pi \, D_b^2$ and a 
volume $\mathcal{V}_b = \frac{\pi D_b^3}{6}$, with $D_b$ being the \textcolor{black}{Sauter mean} diameter, the specific interfacial area $a_\text{w}$ in Eqns. (\ref{EqGas}) and (\ref{EqGas1}) can be written as
\begin{equation}
 a_\text{w} = \frac{A}{\mathcal{V}_\text{w}} = \frac{n \, A_b}{\mathcal{V}_\text{w}}.
\label{Eq4}
\end{equation}

Next, we define the time-averaged volumetric air concentration $C$, which can be understood as the volume of air bubbles per unit volume of air-water mixture, obtained during a measurement with a time scale significantly larger than the time scale of turbulent processes
\begin{equation}
C = \frac{\mathcal{V}_\text{a}}{\mathcal{V}_\text{m}} = \frac{\mathcal{V}_\text{a}}{\mathcal{V}_\text{w} +  \mathcal{V}_\text{a}}, 
\label{Eq5}
\end{equation}
where $\mathcal{V}_\text{m}$ is the mixture volume, and $\mathcal{V}_\text{a}$ is the volume of air, i.e., $\mathcal{V}_a = n \, \mathcal{V}_b$. Solving Eq. (\ref{Eq5}) for $\mathcal{V}_w$ and substitution into Eq. (\ref{Eq4}) yields

\begin{equation}
 a_\text{w} = \frac{n \, A_b}{\mathcal{V}_\text{w}} = \frac{C \,  n \, A_b}{\mathcal{V}_\text{a} \left(1 - C \right)} = \frac{6 \, C \,  n \, \pi \, D_b^2}{n \, \pi \, D_b^3  \left(1 - C \right)} = \frac{6 \,C }{D_b \left(1 - C \right)}.
\label{Eq6}
\end{equation}

When deploying a phase detection intrusive probe, bubbles are not necessarily pierced at their centreline, and a chord length is measured instead of a bubble diameter. The mean air chord length $L_\text{ch}$ can be expressed as \citep{Toombes2002}
\begin{equation}
L_\text{ch} = \frac{C \, V}{F},
\label{Eq7}
\end{equation}
where $V$ is the time-averaged interfacial velocity, and $F$ is the interfacial frequency or bubble count rate. It is known that the bubble diameter is typically larger than the chord length, and a theoretical value of 1.5 has been reported for the ratio of bubble diameter to chord length, i.e., $D_b/L_\text{ch} = 1.5$ \citep{LIU19931061,RUDISULI20121}. Inserting this condition into Eq. (\ref{Eq7}) gives 
\begin{equation}
D_b = \frac{1.5 \, C \, V}{F}.
\label{Eq8}
\end{equation}

In a last step, Eq. (\ref{Eq6}) is combined with Eq. (\ref{Eq8}), leading to the following expression for the specific interfacial area per unit volume of water
\begin{equation}
 a_\text{w} = \frac{6 \, C }{D_b \left(1 - C \right)} = \frac{4 \, F }{V \left(1 - C \right)}. 
 \label{Eq9}
\end{equation}
It is noted that several assumptions were made in the derivation of Eq. (\ref{Eq9}), including (i) \textcolor{black}{an accurate representation of the interfacial area by the Sauter mean diameter}, (ii) spherical bubble shape, and (iii) ratio bubble diameter to chord length $D_b/L_\text{ch} = 1.5$. Note that Eq. (\ref{Eq1}) can be derived in a similar fashion, by defining $a_\text{m} = A/\mathcal{V}_\text{m} = (n \, A_b)/\mathcal{V}_\text{m}$. In order to account for more complex bubble shapes, we define a shape factor $\beta = (A_b \, L_\text{ch})/\mathcal{V}_b$, leading to the following expression for $a_\text{w}$
\begin{equation}
 a_\text{w}  = \beta \, \frac{F}{V \left(1 - C \right)}.
 \label{Eq11}
\end{equation}

Equation (\ref{Eq11}) is in accordance with \cite{Toombes2002}, who stated  that ``\textit{the specific interface area is simply proportional to the number of air-water interfaces}'' within an air-water mixture. An evaluation of the shape factor for spherical bubbles yields $\beta = 4$, demonstrating that Eq. (\ref{Eq11}) can be reconciled with Eq. (\ref{Eq9}). 

\textcolor{black}{Next, it is worthwhile to assess the validity of the derived formulation  with respect to different flow layers, comprising the bubbly flow layer, yet the intermediate flow layer, and the droplet flow layer \citep{Chanson2002c}. 
It is stipulated that Eq. (\ref{Eq9}) holds for bubbly flow with $C < 0.3$, but also for droplet flow with $C > 0.7$, shown in Appendix \ref{App1}. For the intermediate flow layer with $0.3 \leq C \leq 0.7$, the specific interface area simply becomes proportional to the number of air-water interfaces, and is described by Eq. (\ref{Eq11}). In the absence of other evidence, and without the ability to differentiate between air bubbles, waves, and water droplets, a value of $\beta = 4$ is used herein across the entire air-water flow column, implying that Eq. (\ref{Eq9}) should be regarded as a rough approximation of the specific interfacial area.}

\section{Results - Application to stepped spillway flow}
To show the differences between Eq. (\ref{Eq1}) and Eq. (\ref{Eq9}), previously recorded transition flow data from \cite{Kramer2018Transition} were re-analysed, comprising a bed-normal profile of air-water flow properties, measured at step edge 8; the specific flow rate was  $q = 0.067$ m$^2$/s and the chute angle was $\theta = 45^\circ$. For more information about the experimental setup and flow measurement instrumentation, the reader is referred to \cite{Kramer2018Transition}.

\begin{figure}[h!]
\centering
\includegraphics{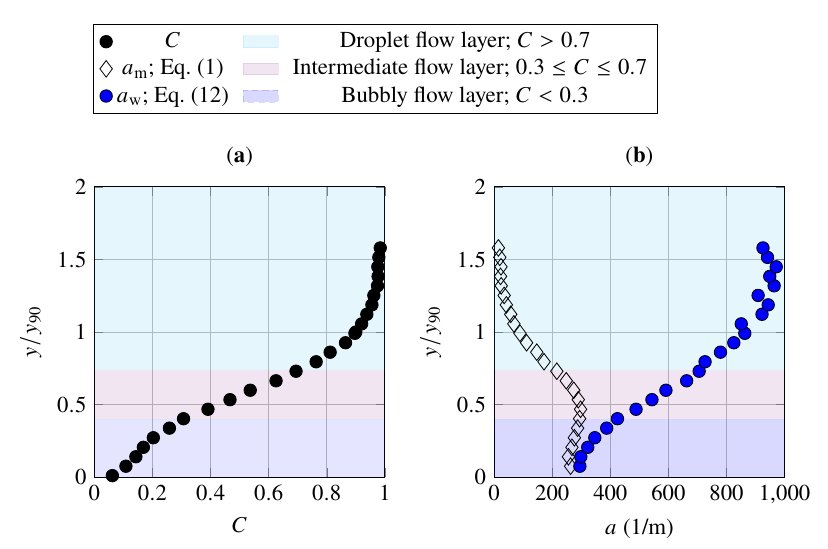}
\caption{Air-water flow properties of flows down a stepped spillway; measurements taken at step edge 8 for $q= 0.067$ m$^2$/s; re-analysis of data from \protect \cite{Kramer2018Transition}: (\textbf{a}) Air concentration distribution; (\textbf{b}) Specific interfacial areas $a_\text{m}$ [Eq. (\ref{Eq1})] and $a_\text{w}$ [Eq. (\ref{Eq9})]}
\label{fig:interfacial}
\end{figure}

Figure \ref{fig:interfacial} shows the results of this re-analysis, including a typical S-shaped air concentration profile (Fig. \ref{fig:interfacial}\textbf{a}), as well as profiles of the specific interfacial areas (Fig. \ref{fig:interfacial}\textbf{b}). Note that the vertical axes were normalised with $y_{90} = y(C = 0.9)$. Considering the data presented in Fig. \ref{fig:interfacial}\textbf{b}, it 
becomes clear that there are large differences in the computation of $a$. For the recorded profiles, the depth-averaged specific interfacial area, defined as \citep{Toombes2000,Toombes2005}
\begin{equation}
\langle a \rangle  = \frac{1}{y_{90}}\int_{y=0}^{y_{90}} a \, \text{d}y, 
\label{adepth}
\end{equation}
yielded values of $\langle a_m \rangle = 158.2$ m$^{-1}$ and $\langle a_w \rangle = 693.2$ m$^{-1}$, respectively. This finding implies that aeration efficiencies, estimated using $a_\text{m}$ instead of $a_\text{w}$, may carry some significant error. \textcolor{black}{It is  further noted that the droplet flow layer ($C > 0.7$) provides a non-negligible contribution to the air-water interface $a_\text{w}$, specifically for $y > y_{90}$  (Fig. \ref{fig:interfacial}\textbf{b}). Therefore, in context of air-water mass transfer calculations, phase-detection measurements need to be extended beyond $y_{90}$, and the upper integration limit of Eq. (\ref{adepth}) needs to be revisited.}

\section{Discussion and Conclusion}
In this work, a formulation for the estimation of  specific interfacial areas in aerated flows is derived, which relates the interfacial area to the volume of water, on the basis of phase-detection intrusive measurements. The underlying assumptions and limitations of this new formulation are discussed, followed by the introduction of a shape factor, which allows application beyond spherical particle shapes. \textcolor{black}{The presented approach is anticipated to be accurate for bubbly flow ($C < 0.3$) and for droplet flow ($C > 0.7$), while the estimation is less precise in  the intermediate flow layer ($0.3 \leq C \leq 0.7$), where the specific interface area becomes proportional to the number of air-water interfaces.}

\textcolor{black}{It is further stipulated that estimated interfacial areas provide a means to facilitate air-water mass transfer calculations in aerated flows, and that previous researchers have used an incorrect definition of the specific interfacial area in their computations, which could have led to significant errors in their estimated aeration efficiencies \citep{Toombes2000,Toombes2002,Toombes2005,Felder2015a,Wuthrich2015,Severi2018,Nina2022,Niu2024,KC02012025}. While the presented formulation corrects for this shortcoming, some other limitations are worthwhile to be mentioned}

\begin{figure}[h!]
\begin{center}
\includegraphics[width=0.7\textwidth]{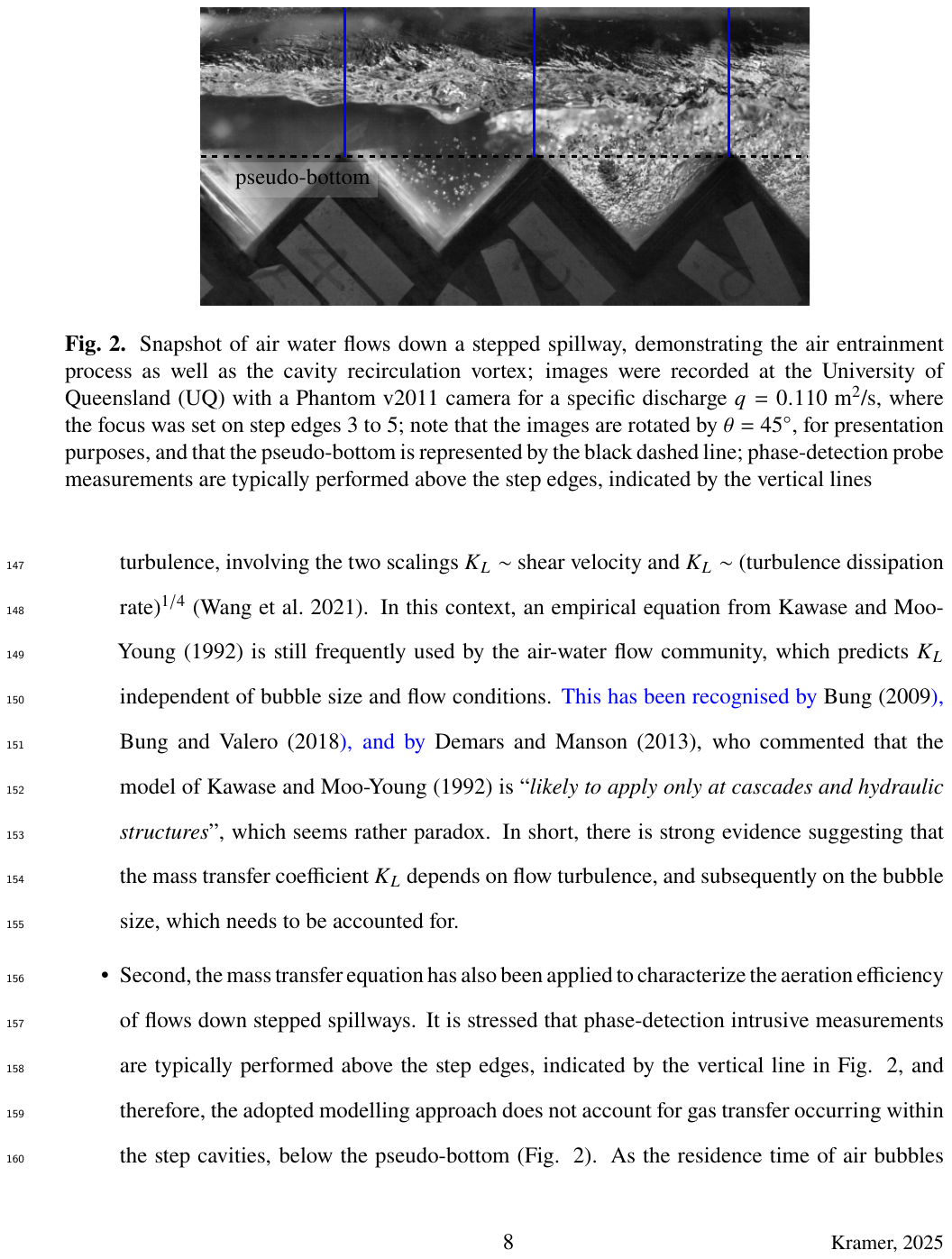}
\end{center}
\caption{Snapshot of air water flows down a stepped spillway, demonstrating the air entrainment process as well as the cavity recirculation vortex; images were recorded at the University of Queensland (UQ) with a Phantom v2011 camera for a specific discharge $q = 0.110$ m$^2$/s, where the focus was set on step edges 3 to 5; note that the images are rotated by $\theta = 45^\circ$, for presentation purposes, and that the pseudo-bottom is represented by the black dashed line; phase-detection probe measurements are typically performed above the step edges, indicated by the vertical lines}
\label{fig2}
\end{figure}

\newpage
\begin{itemize}
\setlength{\itemsep}{0pt}
\item First, it is emphasized that 
stream aeration coefficients $K_L$ are fundamentally controlled by turbulence, involving the two scalings $K_L \sim$  shear velocity and $K_L \sim$  (turbulence dissipation rate)$^{1/4}$ \citep{https://doi.org/10.1029/2020WR028757}. In this context, an empirical equation from \cite{Kawase1992} is still frequently used by the air-water flow community, which predicts $K_L$ independent of bubble size and flow conditions. \textcolor{black}{This has been recognised by \cite{Bung2009}, \cite{Bung2018}, and by} \cite{Demars2013}, who commented that the model of \cite{Kawase1992} is ``\textit{likely to apply only at cascades and hydraulic structures}'', which seems rather paradox. In short, there is strong evidence suggesting that the mass transfer coefficient $K_L$ depends on flow turbulence, and subsequently on the bubble size, which needs to be accounted for. 

\item  Second, the mass transfer equation has also been applied to characterize the aeration efficiency of flows down stepped spillways. It is stressed that phase-detection intrusive measurements are typically performed above the step edges, indicated by the vertical line in Fig. \ref{fig2}, and therefore, the adopted modelling approach does not account for gas transfer occurring within the step cavities, below the pseudo-bottom (Fig. \ref{fig2}). As the residence time of air bubbles trapped inside a cavity vortex may exceed the residence time of air bubbles travelling above the pseudo-bottom by one or two orders of magnitude, the air-water mass transfer within the step cavities may not be negligible. 
\end{itemize}

To summarize, the state-of-the-art approach for estimating aeration efficiencies in flows across hydraulic structures using intrusive phase-detection measurements has some limitations, which have been discussed in detail herein. The newly introduced formulation for the specific interfacial area per unit volume of water provides a key element for a physically based description of gas transfer processes in aerated flows, while other limitations are still to be addressed. 
Altogether, it is hoped that this short contribution provides valuable insights for the future development of advanced air-water mass transfer models for aerated flows.

\section*{Acknowledgements}
The author thanks the School of Civil Engineering (The University of Queensland) for providing full-access to the AEB Hydraulics Laboratory. The exchanges and discussions with Daniel Bung (University of Applied Sciences Aachen), Brian Crookston (Utah State University), John Gulliver (University of Minnesota), and Hubert Chanson (The University of Queensland) are acknowledged, as well as the comments of three anonymous reviewers.

\section*{Data Availability Statement}
No data, models, or code were generated or used during the study.

\input{nomenclature}
\printnomenclature

\bibliographystyle{apacite}
\bibliography{references}

\appendix
\section{Specific interfacial area for droplets}
\label{App1}
\textcolor{black}{Let us consider a droplet flow with $n$ droplets, each droplet having a surface area $A_d = \pi \, D_d^2$ and a volume $\mathcal{V}_d = \frac{\pi D_d^3}{6}$, with $D_d$ being the Sauter mean droplet diameter. The specific interfacial area $a_\text{w}$ can be written as
\begin{equation}
a_\text{w} = \frac{A}{\mathcal{V}_\text{w}} = \frac{6 \, n \, \pi D_d^2}{n \, \pi \, D_d^3} = \frac{6  }{D_d},
\label{Eq15}
\end{equation}
where $n$ is the number of droplets. In accordance with Eq. (\ref{Eq8}), the droplet diameter can be expressed as 
\begin{equation}
D_d = \frac{1.5 \, (1 - C) \, V}{F},
\label{Eq16}
\end{equation}
where the factor $C$ was replaced by $(1-C)$, to account for the chord length of the water phase. Note that for droplet flow with $C > 0.7$, the interfacial frequency $F$ represents a droplet count rate. Combining Eq. (\ref{Eq15}) with Eq. (\ref{Eq16}), the following expression for the specific interfacial area for droplets is obtained
\begin{equation}
 a_\text{w} = \frac{6 }{D_d} = \frac{6 \, F}{1.5 \, (1 - C) \, V} = \frac{4 \, F}{(1 - C) \, V}, 
\end{equation}
which is identical to Eq. (\ref{Eq9}).}

\end{document}

%% file: nomenclature.tex
\nomenclature[A, 01]{\(A\)}{Interfacial area (m$^2$)}
\nomenclature[A, 02]{\(A_b\)}{Air bubble surface area  (m$^2$)}
\nomenclature[A, 02]{\(A_d\)}{\textcolor{black}{Water droplet surface area}  (m$^2$)}
\nomenclature[A, 03]{\(a_\text{m}\)}{Specific interfacial area per unit volume of air-water mixture (1/m)}
\nomenclature[A, 04]{\(a_\text{w}\)}{Specific interfacial area per unit water volume (1/m)}
\nomenclature[A, 05]{\(\langle a \rangle\)}{Depth-averaged interfacial area (1/m)}
\nomenclature[A, 06]{\(C\)}{Time-averaged volumetric air concentration (-)}
\nomenclature[A, 07]{\(c_\text{sat}\)}{Equilibrium gas concentration (kg/m$^3$)}
\nomenclature[A, 08]{\(c_\text{gas}\)}{Dissolved gas concentration in bulk solution (kg/m$^3$)}
\nomenclature[A, 09]{\(D_b\)}{Sauter mean diameter for air bubbles  (m)}
\nomenclature[A, 09]{\(D_d\)}{\textcolor{black}{Sauter mean diameter for water droplets}  (m)}
\nomenclature[A, 10]{\(F\)}{Interfacial frequency, defined as the rate of detected interfaces divided by two (1/s)}
\nomenclature[A, 11]{\(K_L\)}{Mass transfer coefficient (m/s)}
\nomenclature[A, 12]{\(L_\text{ch}\)}{Air chord length (m)}
\nomenclature[A, 13]{\(N\)}{Mass flux through interface (kg/(m$^2$ s)}
\nomenclature[A, 14]{\(n\)}{Number of air bubbles/\textcolor{black}{water droplets}  (-)}
\nomenclature[A, 15]{\(q\)}{Specific water discharge (m$^2$/s)}
\nomenclature[A, 16]{\(s\)}{Distance (m)}
\nomenclature[A, 17]{\(t\)}{Time (s)}
\nomenclature[A, 18]{\(V\)}{Time-averaged interfacial velocity (m/s)}
\nomenclature[A, 19]{\(\mathcal{V}_\text{a}\)}{Volume of air (m$^3$)}
\nomenclature[A, 20]{\(\mathcal{V}_b\)}{Air bubble volume  (m$^3$)}
\nomenclature[A, 20]{\(\mathcal{V}_d\)}{Water droplet volume  (m$^3$)}
\nomenclature[A, 21]{\(\mathcal{V}_\text{m}\)}{Volume of air-water mixture (m$^3$)}
\nomenclature[A, 22]{\(\mathcal{V}_\text{w}\)}{Volume of water (m$^3$)}
\nomenclature[A, 23]{\(y\)}{Direction normal to pseudo-bottom (m)}
\nomenclature[A, 24]{\(y_{90}\)}{Mixture flow depth where $C = 0.9$ (m)}

\nomenclature{\(\beta\)}{Shape factor (-)}
\nomenclature{\(\theta\)}{Chute angle ($^\circ$)}

\renewcommand{\nomgroup}[1]{%
  \setlength\itemsep{0.5em}
  }